\documentclass[conference,a4paper]{IEEEtran}
\IEEEoverridecommandlockouts

\def\Figs{./figs/} % call figures (eps files) if needed

\usepackage[utf8]{inputenc} 
\usepackage[T1]{fontenc}
\usepackage{url}              % provides \url{...}
\usepackage{ifthen}           % provides \ifthenelse

\usepackage[cmex10]{amsmath}  % Use the [cmex10] option to ensure complicance
\usepackage{mleftright}       % fix to wrong spacing of \left-,
\mleftright                   % \middle- \right-commands 

\usepackage{graphicx}         % provides \includegraphics{...} to
\usepackage{booktabs}         % fixes poor spacing in tables and
\usepackage{comment}
\usepackage[url,hyperrefblack,notheorems,IEEEtran]{research17} % customized LaTeX commands
\usepackage{amsmath,amssymb,amsfonts,amsbsy}
\usepackage[colorinlistoftodos]{todonotes} % add inline or margin comments

\usepackage{amsthm} % Note that amsthm must be loaded manually AFTER option *notheorems! 
\newtheorem{theorem}{\mytheoremname}

\newtheorem{definition}{\mydefinitionname}
\newtheorem{remark}{\myremarkname}
\newtheorem{example}{\myexamplename}
\usepackage[capitalize]{cleveref}
\allowdisplaybreaks
\usepackage{array}
\usepackage{multirow}
\usepackage{enumitem} % required for \begin{description}
\usepackage{nicematrix} % for adding vertical lines in a matrix
\renewcommand{\vmat}[1]{\bm{\mat{#1}}} % random matrices
\newcommand{\collect}[1]{\mathscr{#1}} % the collection notation
\newcommand*{\Scale}[2][4]{\scalebox{#1}{\ensuremath{#2}}} % scale equation
\definecolor{colorgreen}{HTML}{009688} % green

\begin{document}

\title{Weakly-Private Information Retrieval From\\ MDS-Coded Distributed Storage} 

%%%%%%
%\author{%
%  \IEEEauthorblockN{Anonymous Authors}
%  \IEEEauthorblockA{%
 %   Please do NOT provide authors' names and affiliations\\
 %   in the paper submitted for review, but keep this placeholder.\\
 %   ISIT23 follows a \textbf{double-blind reviewing policy}.}
%}

%%%%%% Please only add the author names and affiliations for the FINAL
%%%%%% version of the paper, but NOT for the paper submitted for review!
%
%%%%%
%%%%% Single author, or several authors with same affiliation:
 \author{%
   \IEEEauthorblockN{Asbj{\o}rn~O.~Orvedal, Hsuan-Yin~Lin, and Eirik~Rosnes}
   \IEEEauthorblockA{Simula UiB, 
                     N--5006 Bergen, Norway\\
                     Emails:  asbjorn.orvedal@gmail.com, \{lin, eirikrosnes\}{@}simula.no}
                   }

\maketitle

%%%%%
%% Abstract: 
%% If your paper is eligible for the student paper award, please add
%% the comment "THIS PAPER IS ELIGIBLE FOR THE STUDENT PAPER
%% AWARD." as a first line in the abstract. 
%% For the final version of the accepted paper, please do not forget
%% to remove this comment!
%%
\begin{abstract}
We consider the problem of weakly-private information retrieval (WPIR) when data is encoded by a maximum distance separable  code and stored across multiple servers. In WPIR, a user wishes to retrieve a piece of data from a set of servers without leaking too much information about which piece of data she is interested in. We study and provide the first WPIR protocols for this scenario and present results on their optimal trade-off between download rate and information leakage using the maximal leakage privacy metric.
\end{abstract}

\section{Introduction}
\label{sec:intro}

Private information retrieval (PIR),  introduced in a seminal paper by Chor \emph{et al.} \cite{ChorGoldreichKushilevitzSudan95_1, ChorGoldreichKushilevitzSudan98_1},  has been extensively studied for more than two decades in both  the computer science and information theory communities, see, e.g., \cite{BeimelIshaiKushilevitRaymond02_1, Yekhanin10_1, CorriganGibbsKogan20_1, ChanHoYamamoto15_1, Freij-HollantiGnilkeHollantiKarpuk17_1, TajeddineGnilkeElRouayheb18_1} and references therein. In PIR, the objective is to download a piece of data stored on a set of servers without leaking any information about which piece of data is being requested to the servers storing the data, while minimizing the overall communication cost. As the upload cost is typically much lower than the download cost, the download rate, defined as the ratio between the amount of requested information and the amount of downloaded information, is used as a measure to compare different PIR protocols. When data is replicated across several servers, the maximum achievable download rate, referred to as the PIR capacity, was derived in \cite{SunJafar17_1}, while the capacity for the case where the data is encoded by a maximum distance separable (MDS) code and stored across a set of servers was settled in \cite{BanawanUlukus18_1}. Arbitrary linear storage codes were considered in \cite{KumarLinRosnesGraellAmat19_1, LinKumarRosnesGraellAmat18_2}.

Weakly-private information retrieval (WPIR), introduced independently by Lin \emph{et al.} \cite{LinKumarRosnesGraellAmatYaakobi19_1} and Samy \emph{et al.} \cite{SamyTandonLazos19_1}, is a relaxed version of PIR that allows for reducing the download cost at the expense of some information leakage on the identity of the requested piece of data to the servers storing it. So far, only the case of replicated data (across servers) and the single server case have been considered in the literature \cite{LinKumarRosnesGraellAmatYaakobi21_1, LinKumarRosnesGraellAmatYaakobi22_1, QianZhouTianLiu22_1, SamyAttiaTandonLazos21_1, ZhouGuoTian20_1, YakimenkaLinRosnesKliewer22_1, WengYakimenkaLinRosnesKliewer22_1}, while in this work we consider for the first time the case where the data is encoded by an MDS code and stored across multiple servers. WPIR protocols allow for a trade-off between download rate and privacy leakage, and the optimal trade-off curve for the case of multiple servers is still an open problem. % even for the case of replicated data. 
As in previous works, we consider the maximal leakage (MaxL) privacy metric~\cite{Smith09_1, BartheKopf11_1, IssaWagnerKamath20_1}. Our main contributions are as follows.

\begin{itemize}
\item We adapt the PIR protocols in \cite{ZhuYanQiTang20_1, ZhouTianSunLiu20_1} for MDS-coded databases to allow for information leakage. The adapted protocols from \cite{ZhuYanQiTang20_1, ZhouTianSunLiu20_1}, referred to as the ZYQT and ZTSL MDS-WPIR schemes, respectively, yield a trade-off between download rate and information leakage, and we show that for the MaxL privacy metric the optimal trade-off is the solution of a convex optimization problem (see Theorem~\ref{thm:rate-leakage_convex-optimization}). The optimized ZYQT MDS-WPIR scheme yields the best trade-off but also has the largest query space.

% provides a trade-off between download rate and information leakage, using the MaxL privacy metric, as the solution of a convex optimization problem, and the one providing the best trade-off depends on the scenario considered.

\item We propose a \emph{new} WPIR protocol, referred to as the OLR MDS-WPIR scheme, with a much smaller query space than the ZYQT scheme while providing an equally good or better trade-off between download rate and information leakage. As for the ZYQT and ZTSL MDS-WPIR schemes, the optimal trade-off is the solution of a convex optimization problem (see Theorem~\ref{thm:rate-leakage_convex-optimization}).

%\item We propose two WPIR protocols for the case of MDS-coded data stored across multiple servers based, respectively, on the PIR protocols in \cite{ZhuYanQiTang20_1, ZhouTianSunLiu20_1}. Both protocols 

%\item For the second proposed protocol, we provide an analytical expression for the optimal rate-leakage trade-off curve. 

\end{itemize}

\section{Preliminaries and System Model}
\label{sec:preliminaries-SystemModel}

\subsection{Notation}% xxx
\label{sec:notation}

We denote by $\Naturals$ the set of all positive integers, and $[a:b]\eqdef\{a,a+1,\ldots,b\}$ for $a,b\in\{0\}\cup\Naturals$, $a \leq b$. Vectors (normally row-wise) are denoted by bold letters, random variables (RVs) (either scalar or vector) by uppercase letters, and sets by calligraphic uppercase letters, e.g., $\vect{x}$, $X$, and $\set{X}$, respectively. Matrices are denoted by sans serif letters, while random matrices are represented by bold sans serif capital letters, e.g., $\vmat{X}$, and $\mat{x}$ represents its realization. The all-one (all-zero) row vector is denoted by $\vect{1}$ ($\vect{0}$), and its length will be clear from the context. When a set of indices $\set{S}$ is given, $\vect{X}_{\set{S}}$ denotes $\{\vect{X}_s\colon s\in\set{S}\}$. $\E[X]{\cdot}$ denotes  expectation with respect to the RV $X$. $X\sim P_X$ denotes an RV distributed according to a probability mass function (PMF) $P_X(x)$, $x\in\set{X}$, and $X\sim\Uniform{\set{S}}$ a uniformly-distributed RV over a set $\set{S}$. $\HP{\cdot}$ denotes the entropy function,  
$\trans{(\cdot)}$ the transpose of a matrix, and $\gcd(a,b)$ the greatest common divisor  of two positive integers $a$ and $b$.

\subsection{System Model}
\label{subseq:system-model}

We consider an MDS-coded distributed storage system (DSS) with $\const{N}$ noncolluding servers that store $\const{M}$ independent files $\vmat{W}^{(1)},\ldots,\vmat{W}^{(\const{M})}$, where each file is represented as a random matrix $\vmat{W}^{(m)}=\bigl(W_{i,j}^{(m)}\bigr)$ of size $\lambda\times\const{K}$, $\lambda,\const{K}\in\Naturals$. Each file $\vmat{W}^{(m)}$ is encoded row-wise using an $[\const{N},\const{K}]$ MDS code $\set{C}$ over some finite field $\Field_q$ of size $q \geq \const{N}$ resulting in the codewords $\bigl(X^{(m)}_{i,1},\ldots,X^{(m)}_{i,\const{N}}\bigr)=(W^{(m)}_{i,1},\ldots,W^{(m)}_{i,\const{K}})\mat{G}^{\set{C}}$, $i\in [0:\lambda-1]$, where $\mat{G}^{\set{C}}$ denotes a generator matrix for $\set{C}$. Denote by $\vect{X}^{(m)}_j\eqdef\trans{\bigl(X^{(m)}_{0,j},\ldots,X^{(m)}_{\lambda-1,j}\bigr)}$ a vector consisting of $\lambda$ code symbols generated by the code $\set{C}$. Then, the $j$-th server stores $\vect{X}_j\eqdef\trans{\bigl(\trans{(\vect{X}^{(1)}_j)}\vert\cdots\vert\trans{(\vect{X}^{(\const{M})}_j)}\bigr)}$, $j\in[1:\const{N}]$.

To retrieve a file $\vmat{W}^{(M)}$, from the MDS-coded DSS, the user sends a query $\vmat{Q}_j$ to the $j$-th server for all $j\in[1:\const{N}]$. Here, $M\sim\Uniform{[1:\const{M}]}$ is an  RV representing the desired file index. In response to the received query, server $j$ returns the answer $\vmat{A}_j$, which is a function of $\vmat{Q}_j$ and the code symbols $\vect{X}_j$ stored in the server, back to the user.  We formally describe an MDS-coded $(\const{M},\const{N},\const{K})$ WPIR scheme as follows.

\begin{definition}[MDS-WPIR Scheme]
  \label{def:def_MNK-WPIRscheme}
  An $(\const{M},\const{N},\const{K})$ MDS-WPIR scheme for an $[\const{N},\const{K}]$ MDS-coded DSS with $\const{N}$ noncolluding servers consists of:
  \begin{itemize}
  \item $\const{M}$ independent files $\vmat{W}^{(m)}$ of size $\lambda\times\const{K}$, for some $\lambda\in\Naturals$, $m \in [1:\const{M}]$.
    
  \item A global random strategy $\vmat{S}$, whose alphabet is $\set{S}$. In general, the realization of $\vmat{S}$ is a matrix.
    
  \item An $(\const{N},\const{K})$ MDS storage code $\set{C}$ that encodes the file $\vmat{W}^{(m)}$ into the matrix $\vmat{X}^{(m)}=\bigl(\vect{X}^{(m)}_1\vert\cdots\vert\vect{X}^{(m)}_\const{N}\bigr)$ as described above, $m\in[1:\const{M}]$.    
    
  \item $\const{N}$ queries $\vmat{Q}_j=\phi_j(M,\vmat{S})$ with alphabet $\set{Q}_j$, $j\in[1:\const{N}]$, that are generated by the query-encoding functions $\phi_j$.  Query $\vmat{Q}_j$ is sent to the $j$-th server.
    
  \item $\const{N}$ answers $\vmat{A}_j=\psi_j(\vmat{Q}_j,\vect{X}_j)$ with alphabet $\set{A}=\Field_q$, $j\in [1:\const{N}]$, that are constructed by the answer functions $\psi_j$. All answers $\vmat{A}_j$ are sent back to the user.

  \item $\const{N}$ answer lengths $\ell_j(\vmat{Q}_j)\in\{0\}\cup\Naturals$, $j\in [1:\const{N}]$,  each being a function of the corresponding query $\vmat{Q}_j$.
  \end{itemize}

  In addition, the scheme should satisfy the following condition of perfect retrievability:
  \begin{IEEEeqnarray*}{c}
    \bigHPcond{\vmat{W}^{(M)}}{\vmat{A}_{[1:\const{N}]},\vmat{Q}_{[1:\const{N}]},M}=0.
    %\label{eq:retrievability}
  \end{IEEEeqnarray*}
\end{definition}

\subsection{Maximal Leakage Metric}
\label{maximal-leakage-metric}

From Definition~\ref{def:def_MNK-WPIRscheme}, one can notice that at the $j$-th server, the requested file index $M$ can be inferred by observing the query distribution $P_{\vmat{Q}_j}$, which results in an information leakage on $M$ to the servers. In this work, we adopt a meaningful information-theoretic privacy metric from the computer science literature, the MaxL metric, to measure information leakage. Formally, given the query distributions $P_{M,\vmat{Q}_j}$, $j\in [1:\const{N}]$, of a given $(\const{M},\const{N},\const{K})$ WPIR scheme $\code{C}$, the overall MaxL about $M$ of $\code{C}$ is defined as
\begin{IEEEeqnarray*}{c}
  \rho^{(\textnormal{MaxL})}(\collect{C})\eqdef\max_{j\in [1:\const{N}]}\ML{M}{\vmat{Q}_j},
\end{IEEEeqnarray*}
where
\begin{IEEEeqnarray*}{c}
  \ML{M}{\vmat{Q}}\eqdef\log_2{\Biggl(\sum_{\mat{q}\in\set{Q}}\max_{m\in[\const{M}]}P_{\vmat{Q}|M}(\mat{q}|m)\Biggr)}.
\end{IEEEeqnarray*}

Note that an $[\const{N},\const{K}]$ MDS-coded PIR scheme is an $(\const{M},\const{N},\const{K})$ WPIR scheme $\code{C}$ that satisfies  $\rho^{(\textnormal{MaxL})}(\code{C})=0$, such a condition is refereed to as the \emph{perfect privacy} constraint.

\subsection{WPIR Download Cost and Rate}
\label{sec:WPIR-download-cost-and-rate}

The overall download cost (in number of symbols over $\Field_q$) and rate of a WPIR scheme $\code{C}$, denoted by $\const{D}(\code{C})$ and $\const{R}(\code{C})$, respectively, are given by
\begin{IEEEeqnarray*}{c}
  \const{D}(\collect{C})=\sum_{j=1}^{\const{N}}\E[\vmat{Q}_j]{\ell_j(\vmat{Q}_j)} \textnormal{ and }\const{R}(\collect{C})\eqdef\frac{\lambda\const{K}}{\const{D}(\collect{C})}.
\end{IEEEeqnarray*}

\section{General MDS-WPIR Schemes}
\label{sec:general-MDS-WPIRschemes}

In this section, we give a general description of the $(\const{M},\const{N},\const{K})$ MDS-WPIR schemes we consider in this work. We start by reviewing two MDS-PIR capacity-achieving schemes for small file sizes, namely the ZYQT scheme~\cite{ZhuYanQiTang20_1} and the ZTSL scheme~\cite{ZhouTianSunLiu20_1}.\footnote{Precisely, the ZTSL scheme we consider here is the so-called Construction-A MDS-PIR code that is referred in~\cite[Sec.~III]{ZhouTianSunLiu20_1}.}

\subsection{The ZYQT Scheme and the ZTSL Scheme}
\label{sec:ZYQT-and-ZTSL-scheme}

\subsubsection{Storage Data Structure}
\label{sec:storage-data-structure}

The following \emph{effective code parameters} are universally defined for an MDS-coded DSS:
\begin{IEEEeqnarray*}{c}
  n\eqdef\frac{\const{N}}{\gcd(\const{N},\const{K})},\quad k \eqdef \frac{\const{K}}{\gcd(\const{N},\const{K})},\quad r\eqdef n-k.
\end{IEEEeqnarray*}
Moreover, the subpacketization size for each file is given by $\lambda=n-k$. For ease of exposition, we further append $k$ dummy variables $X^{(m)}_{i,j}\equiv 0$ for $i\in [n-k:n-1]$, $j\in[1:\const{N}]$, such that for all $m\in [1:\const{M}]$,
\setlength{\arraycolsep}{0.3pt}
\begin{IEEEeqnarray}{c}
  \vmat{X}^{(m)}=
  \begin{pNiceMatrix}% [nullify-dots]
    X^{(m)}_{0,1}&  X^{(m)}_{0,2} & \Cdots &  X^{(m)}_{0,\const{N}}
    \\
    \Vdots &\Vdots & \Ddots &\Vdots
    \\
    X^{(m)}_{n-k-1,1}&  X^{(m)}_{n-k-1,2} & \Cdots &  X^{(m)}_{n-k-1,\const{N}}
    \\
    0 & 0 & \Cdots & 0      
    \\
    \Vdots & \Vdots& \ddots & \Vdots
    \\
    0 & 0 & \Cdots & 0
    \CodeAfter
    \SubMatrix.{4-4}{6-4}\}[right-xshift=0.5em,name=A]
    \tikz \node [right] at (A-right.east) {$k$ rows};
  \end{pNiceMatrix}\,.\IEEEeqnarraynumspace\label{eq:def_X-data}
\end{IEEEeqnarray}

\subsubsection{Query Generation}
\label{sec:query-generation}

The query generation is the main difference among the $(\const{M},\const{N},\const{K})$ MDS-WPIR schemes. In our context, we will make use of the set
\begin{IEEEeqnarray*}{rCl}
  \set{P}^n_k& \eqdef &\bigl\{
    \trans{\vect{s}}=\trans{(s_1,\ldots,s_k)}\colon s_i,s_{i'}\in [0:n-1],
    \nonumber\\
    &&\hspace*{3.0cm}\> s_i\neq s_{i'},\,\forall\,i, i' \in [1:k], i\neq i'\bigr\}\IEEEeqnarraynumspace
\end{IEEEeqnarray*}
of column vectors. 
The global random strategy alphabet for the ZYQT and ZTSL schemes are, respectively, given by
\begin{IEEEeqnarray*}{rCl}
  \set{S}_\textnormal{ZYQT}& \eqdef &\{\mat{s}=(\trans{\vect{s}_1},\ldots,\trans{\vect{s}_\const{M}})\colon\trans{\vect{s}_{m'}}\in\set{P}^n_k,\,m'\in [1:\const{M}]\},\IEEEeqnarraynumspace
  \\[1mm]
  \set{S}_\textnormal{ZTSL}& \eqdef &\left\{\vect{s}\in [0:n-1]^{\const{M}}\colon\left(\sum_{m'=1}^\const{M}s_{m'}\right)\bmod n = 0\right\}.\IEEEeqnarraynumspace
\end{IEEEeqnarray*}
Note that $\card{\set{S}_\textnormal{ZYQT}}=\bigl(\binom{n}{k} k!\bigr)^{\const{M}}$ and $\card{\set{S}_\textnormal{ZTSL}}=n^{\const{M}-1}$. Since the cost of uploading the queries for an MDS-PIR scheme depends on the cardinality of the global random strategy alphabet, it is apparent that the ZTSL scheme has a lower upload cost than the ZYQT scheme. It is also worth mentioning that MDS-PIR schemes are generally constructed using an $\vmat{S}$ that is uniformly distributed over the set $\set{S}$.

We next present the original query generation for the ZYQT and ZTSL MDS-PIR schemes for retrieving the $m$-th file $\vmat{X}^{(m)}$, $m\in [1:\const{M}]$. Notice that we do not adopt the uniformly-distributed $\vmat{S}$ here. Thus, the leakage $\rho^{(\textnormal{MaxL})}$ is not necessarily equal to $0$. We refer to the corresponding proposed schemes as the ZYQT MDS-WPIR and  ZTSL MDS-WPIR schemes and denote them by $\collect{C}_\textnormal{ZYQT}$ and $\collect{C}_\textnormal{ZTSL}$, respectively.

\begin{description}
\item[$\code{C}_\textnormal{ZYQT}:$] The query $\mat{q}_j\in\set{Q}_j$, $j\in [1:\const{N}]$, generated from the query-encoding function $\phi_j$
is defined as
  \begin{IEEEeqnarray*}{rCl}
    \mat{q}_j& = &(\trans{\vect{s}_1},\ldots,\trans{\vect{s}}_{m-1},\bigl(\trans{\vect{s}_m}+(j-1)\trans{\vect{1}}\bigr)\bmod n,\nonumber\\
    &&\quad \>\trans{\vect{s}_{m+1}},\ldots,\trans{\vect{s}_\const{M}}),\quad\trans{\vect{s}_m}\in\set{P}^n_k,\, m\in [1:\const{M}].\IEEEeqnarraynumspace\label{eq:queries_ZYQT-WPIRscheme}
  \end{IEEEeqnarray*}
  
\item[$\code{C}_\textnormal{ZTSL}:$] The query $\mat{q}_j\in\set{Q}_j$, $j\in [1:\const{N}]$, is generated by
  \setlength{\arraycolsep}{0.60pt}
  \begin{IEEEeqnarray*}{rCl}
    \mat{q}_j& = &\left[
      \begin{pNiceArray}{ccccccc}
        s_1& \Cdots & s_{m-1} & (s_{m}+(j-1)) & s_{m+1} &\Cdots & s_{\const{M}}
        \\
        \Vdots & \Ddots &\Vdots &\Vdots &\Vdots &\Ddots &\Vdots
        \\
        s_1& \Cdots & s_{m-1} & (s_{m}+(j-1)) & s_{m+1} &\Cdots & s_{\const{M}}
        \CodeAfter
        \tikz
        % Simple brace
        \draw [decorate, decoration = {brace, amplitude=5pt},very thick] (-0.1,0.75) -- (-0.1,-0.65) node[black,midway,xshift=0.60cm] {\scriptsize $k$ rows};
        % \SubMatrix{.}{1-1}{3-1}{\{}[left-xshift=2cm,name=A]
        % \tikz \node [right] at (A-left.east) {$\kappa$ rows};
      \end{pNiceArray}\right.
    \nonumber\\[2mm]
    &&\hspace*{1.85cm}+\>\left.
      \begin{pNiceMatrix}% [nullify-dots]
        0 & 0 & \Cdots & 0
        \\
        1 & 1 & \Cdots & 1
        \\
        \Vdots & \Vdots &\ddots &\Vdots
        \\
        k-1    &k-1     &\Cdots&k-1
        \CodeAfter
        \UnderBrace[yshift=15pt]{1-1}{3-4}{\const{M}\textnormal{ columns}}
      \end{pNiceMatrix}\right]\bmod n,\IEEEeqnarraynumspace\label{eq:queries_ZTSL-WPIRscheme}
    \\[2mm]
  \end{IEEEeqnarray*}
  where $\vect{s}\in\set{S}_\textnormal{ZTSL}$.
\end{description}

\subsubsection{Answer Construction}
\label{sec:answer-construction}

Upon receiving a query (matrix)%\footnote{\eirik{To simplify  notation the dependency on $j$ is omitted.}}
\begin{IEEEeqnarray*}{c}
  \setlength{\arraycolsep}{2pt}
  \mat{q}_j=
  \begin{pNiceMatrix}
    q_{1,1} & q_{1,2} & \Cdots & q_{1,\const{M}}
    \\
    \Vdots  & \Vdots  & \Ddots & \Vdots
    \\
    q_{k,1} & q_{k,2} & \Cdots & q_{k,\const{M}}
  \end{pNiceMatrix},
\end{IEEEeqnarray*}
the $j$-th server uses the  answer function $\psi_j$ to construct the answer
\begin{IEEEeqnarray*}{c}
  \vmat{A}_j=\psi_j(\mat{q}_j,\vect{X}_j)=\trans{\Biggl(\sum_{m'=1}^\const{M}\! X^{(m')}_{q_{1,m'},j},\cdots,\sum_{m'=1}^\const{M}\! X^{(m')}_{q_{k,m'},j}\Biggr)}\IEEEeqnarraynumspace%\label{eq:answers_MDS-WPIRschemes}
\end{IEEEeqnarray*} 
consisting of $k$ sub-responses. With the storage data defined in \eqref{eq:def_X-data}, the length of the answer  is given by the number of nonzero components in $\vmat{A}_j$, which is equal to
\begin{IEEEeqnarray*}{c}
  \ell_j(\mat{q}_j)=\sum_{i=1}^k\indicatorfunction\biggl\{\min_{m'\in [1:\const{M}]}q_{i,m'}\leq n-k-1\biggr\},
  %\label{eq:answer-length_MDS-WPIRscheme}
\end{IEEEeqnarray*}
where $\indicatorfunction\{\textnormal{statement}\}$ is the indicator function whose value is $1$ if the statement is true and $0$ otherwise.

Finally, we remark that according to the query constructions for both the ZYQT and  ZTSL MDS-WPIR schemes, the file $\vmat{W}^{(m)}$ can always be reconstructed by the MDS property of the storage code $\mathcal{C}$ (the so-called $\const{K}$-out-of-$\const{N}$ property).

\subsection{Time-Sharing MDS-WPIR Scheme}
\label{sec:Time-Sharing-MDS-WPIRscheme}

Clearly, selecting a different global random strategy $\vmat{S}$ leads to a different WPIR rate and privacy leakage of an MDS-WPIR scheme. This work aims to achieve the best trade-off between download rate and privacy leakage by using the best $\vmat{S}$ for an MDS-WPIR scheme. However, the minimization problem of the information leakage for a given WPIR rate over the global random strategy for an MDS-WPIR scheme is generally not convex. Hence, in order to easily tackle the optimization problem, we make use of a time-sharing principle to \emph{convexify} the optimization problem for determining the best rate-leakage trade-off~\cite[Sec.~VII]{LinKumarRosnesGraellAmatYaakobi22_1}.

\begin{definition}[Time-Sharing MDS-WPIR Scheme]
  \label{def:time-sharing-MDS-WPIRscheme}
  Consider an MDS-WPIR scheme $\mathring{\code{C}}$ with query-encoding functions $\mathring{\phi_j}$, answer functions $\mathring{\psi_j}$, and a global random strategy $\mathring{\vmat{S}}$. The time-sharing MDS-WPIR scheme of $\mathring{\code{C}}$ is made by the query-encoding functions $\phi_j=\mathring{\phi}_{\sigma^{T-1}(j)}(M,\vmat{S})$ and the answer functions $\psi_j=\mathring{\psi}_{\sigma^{T-1}(j)}\bigl(\mathring{\phi}_{\sigma^{T-1}(j)}(M,\vmat{S}),\vect{X}_j\bigr)$, $j\in [1:\const{N}]$, for a given requested file index $M$, where $T\sim\Uniform{[1:\const{N}]}$, and $\sigma(\cdot)$ denotes a left circular shift, while $l$ left circular shifts are obtained through function composition and denoted by $\sigma^l(\cdot)$. Such an MDS-WPIR scheme $\code{C}$ is called the time-sharing scheme of $\mathring{\code{C}}$.
\end{definition}
\begin{remark}~
\label{rem:properties_time-sharing-MDS-WPIRscheme}
  \begin{itemize}
  \item A time-sharing MDS-WPIR scheme always has equal information leakage at each server~\cite[Th.~1]{LinKumarRosnesGraellAmatYaakobi22_1}.
  \item In the following, unless specified otherwise, all the MDS-WPIR schemes we discuss are assumed to be already post-processed by applying the time-sharing principle, and the minimization of MaxL is also done for the time-sharing scheme of an MDS-WPIR scheme.
  \end{itemize}
\end{remark}

\subsection{Minimization of  MaxL for MDS-WPIR Schemes}
\label{sec:minimization_MaxL_MDS-WPIRschemes}

Denote by $z_{\mat{s}}\eqdef P_{\vmat{S}}(\mat{s})$ the PMF of the random strategy $\vmat{S}$. It can be shown that both the MaxL $\rho^{(\textnormal{MaxL})}(\code{C})$ and the WPIR download cost $\const{D}(\code{C})$ of a given MDS-WPIR scheme $\code{C}$ can be expressed in terms of $z_{\mat{s}}$, $\mat{s}\in\set{S}$. Thus, the minimization of $\rho^{(\textnormal{MaxL})}(\code{C})$ under a download cost constraint $\const{D}(\code{C}) \leq \const{D}$ can be re-written in terms of the variables $\{z_{\mat{s}}\}_{\mat{s}\in\set{S}}$ as the optimization problem
\begin{IEEEeqnarray}{rCl}
  \IEEEyesnumber\label{eq:optimization_MaxL-download}
  \IEEEyessubnumber*
  \textnormal{minimize} & &\qquad\rho^{(\textnormal{MaxL})}(\{z_{\mat{s}}\}_{\mat{s}\in\set{S}})
  \label{eq:objective-ft_leakage}\\
  \textnormal{subject to} & &\qquad \const{D}(\{z_{\mat{s}}\}_{\mat{s}\in\set{S}})\leq\const{D},\label{eq:PMFs_download-constraint}
  \\[1mm]
  & &\qquad\sum_{\mat{s}\in\set{S}}z_{\mat{s}}=1.\label{eq:PMF_random-strategy}
\end{IEEEeqnarray}

The following theorem can be proved using a similar argument as in \cite[Sec.~VII]{LinKumarRosnesGraellAmatYaakobi22_1}.
\begin{theorem}
\label{thm:rate-leakage_convex-optimization}
    The optimization problem~\eqref{eq:optimization_MaxL-download} is convex.
\end{theorem}
All the rate-leakage trade-off curves of the MDS-WPIR schemes we study in this work are based on solving the convex optimization problem above.

\section{New Proposed MDS-WPIR Scheme}
\label{sec:proposed-MDS-WPIRscheme}

This section presents a new MDS-WPIR scheme, referred to as the OLR MDS-WPIR scheme. We first present an example illustrating the motivation for studying the new MDS-WPIR scheme in \cref{sec:WPIR_N3K2M2}. In particular, we will show that the ZTSL MDS-WPIR scheme is naturally not a good scheme as it is not functional in the high-rate region when there is  leakage.

\subsection{Motivating Example: $(\const{M},\const{N},\const{K})=(2,3,2)$}
\label{sec:WPIR_N3K2M2}

For %the case of 
$(\const{N},\const{K})=(3,2)$, we have the effective code parameters
\begin{IEEEeqnarray*}{c}
  n=\frac{\const{N}}{\gcd(\const{N},\const{K})}=3,\, k=\frac{\const{K}}{\gcd(\const{N},\const{K})}=2,\, r=n-k=1,\IEEEeqnarraynumspace
\end{IEEEeqnarray*}
and the subpacketization size for each file is $\lambda=n-k=1$.

For the $(2,3,2)$ ZTSL MDS-WPIR scheme, we have
%\begin{IEEEeqnarray*}{c}
  $\set{S}_\textnormal{ZTSL}=\{(0,0),(1,2),(2,1)\}$,
%\end{IEEEeqnarray*}
and the corresponding conditional query PMF $P_{\vmat{Q}_j\mid M}(\mat{q}_j\mid m)$ and answer lengths are  as follows:
\NiceMatrixOptions{cell-space-limits = 1mm}
\begin{IEEEeqnarray}{c}
  \Scale[0.85]{
    \begin{pNiceMatrix}[first-row,first-col,nullify-dots]
      {\small P_{\vmat{Q}_j|M}(\mat{q}_j\mid m)} & \bigl(
      \begin{smallmatrix}
        0 & 0
        \\
        1 & 1
      \end{smallmatrix}\bigr) & \bigl(
      \begin{smallmatrix}
        1 & 2
        \\
        2 & 0
      \end{smallmatrix}\bigr) & \bigl(
      \begin{smallmatrix}
        2 & 1
        \\
        0 & 2
      \end{smallmatrix}\bigr) & \bigl(
      \begin{smallmatrix}
        1 & 0
        \\
        2 & 1
      \end{smallmatrix}\bigr) & \bigl(
      \begin{smallmatrix}
        2 & 2
        \\
        0 & 0
      \end{smallmatrix}\bigr) & \bigl(
      \begin{smallmatrix}
        0 & 1
        \\
        1 & 2
      \end{smallmatrix}\bigr) & \bigl(
      \begin{smallmatrix}
        2 & 0
        \\
        0 & 1
      \end{smallmatrix}\bigr) & \bigl(
      \begin{smallmatrix}
        0 & 2
        \\
        1 & 0
      \end{smallmatrix}\bigr) & \bigl(
      \begin{smallmatrix}
        1 & 1
        \\
        2 & 2
      \end{smallmatrix}\bigr) 
      \\
      1 & \frac{z_1}{3} & \frac{z_2}{3} & \frac{z_3}{3} & \frac{z_1}{3} & \frac{z_2}{3} & \frac{z_3}{3}
      & \frac{z_1}{3} & \frac{z_2}{3}  & \frac{z_3}{3}
      \\
      2 & \frac{z_1}{3} & \frac{z_2}{3} & \frac{z_3}{3} & \frac{z_2}{3} & \frac{z_3}{3} & \frac{z_1}{3}
      & \frac{z_3}{3} & \frac{z_1}{3}  & \frac{z_2}{3}
      \\*\hline
      P_{\vmat{Q}_j}(\mat{q}_j) & \frac{z_1}{3} & \frac{z_2}{3}  & \frac{z_3}{3} & \frac{z_1+z_2}{6} & \frac{z_2+z_3}{6} & \frac{z_1+z_3}{6}
      & \frac{z_1+z_3}{6} & \frac{z_1+z_2}{6} & \frac{z_2+z_3}{6}
      \\*\hdottedline
      \ell_j(\mat{q}_j) & 1 & 1  & 1 & 1 & 1 & 1 & 2 & 2 & 0
      \CodeAfter
       \SubMatrix\{{1-1}{2-1}.[left-xshift=1.5em,name=B]
      \tikz \node [left] at (B-left.west) {$m$};
    \end{pNiceMatrix}},\nonumber\\*\IEEEeqnarraynumspace\label{eq:PQ_M_ZTSL_N3K2M2}
\end{IEEEeqnarray}
where $z_j\eqdef\Prs{\vect{s}_j}$ for $\vect{s}_j=(j-1,(n-j+1)\bmod n)\in\set{S}_\textnormal{ZTSL}$, $j\in[1:n]$. A simple calculation gives
\begin{IEEEeqnarray*}{c}
  \const{D}(\collect{C}_\textnormal{ZTSL})=3+z_1,\quad 0\leq z_1\leq 1,
\end{IEEEeqnarray*}
which indicates that  $\const{D}(\collect{C}_\textnormal{ZTSL})$ can only range between $3$ and $4$, and never reaches $\const{R}=\nicefrac{\lambda\const{K}}{\const{D}}=\nicefrac{2}{\const{D}}=1$. Thus, the ZTSL MDS-WPIR scheme can not operate in the high-rate region.

\subsection{New $(\const{M},\const{N},\const{K})$ MDS-WPIR Scheme}
\label{sec:new-WPIRscheme}

We now describe the new proposed $(\const{M},\const{N},\const{K})$ MDS-WPIR scheme, referred to as the OLR MDS-WPIR scheme and denoted by $\collect{C}_\textnormal{OLR}$. Here, only the query generation is presented, as its answer construction is the same as Section~\ref{sec:answer-construction}. 

\subsubsection{Query Generation}
\label{sec:query-generation_new-WPIRscheme}

The strategy set for our new MDS-WPIR scheme is defined as
\begin{IEEEeqnarray*}{rCl}
  \set{S}_\textnormal{OLR}\eqdef\Biggl\{\mat{s}& = &(\trans{\vect{s}_1},\ldots,\trans{\vect{s}_{\const{M}-1}})\colon \trans{\vect{s}_{m'}}\in\set{P}^n_k,\,m'\in[1:\const{M}],\nonumber\\
  &&\hspace*{2.5cm}\>\left(\sum_{m'=1}^{\const{M}}\trans{\vect{s}_{m'}}\right)\bmod n =\trans{\vect{0}}\Biggr\}.\IEEEeqnarraynumspace
\end{IEEEeqnarray*}
By definition,  $\card{\set{S}_{\textnormal{OLR}}} \leq \bigl(\binom{n}{k} k!\bigr)^{\const{M}-1} < \card{\set{S}_{\textnormal{ZYQT}}}=\bigl(\binom{n}{k} k!\bigr)^{\const{M}}$, as we do not  include all the possible vectors $\trans{\vect{s}_{m'}}\in\set{P}^n_k$.

The query $\mat{q}_j\in\set{Q}_j$, $j\in [1:\const{N}]$, for retrieving the $m$-th file, $m\in [1:\const{M}]$, is defined as
\begin{IEEEeqnarray}{c}
    \mat{q}_j=(\trans{\vect{s}_1},\ldots,\trans{\vect{s}}_{m-1},\trans{\vect{q}_m},\trans{\vect{s}_{m}},\ldots,\trans{\vect{s}_{\const{M}-1}}),
    \IEEEeqnarraynumspace\label{eq:queries_OLR-WPIRscheme}
\end{IEEEeqnarray}
where $(\trans{\vect{s}_1},\ldots,\trans{\vect{s}_{\const{M}-1}})=\mat{s}\in\set{S}_{\textnormal{OLR}}$ and
\begin{IEEEeqnarray*}{c}
    \trans{\vect{q}}_{m}\eqdef\biggl((j-1)\trans{\vect{1}}-\sum_{m'\in[1:\const{M}-1]}\trans{\vect{s}_{m'}}\biggr)\bmod n.
\end{IEEEeqnarray*}

\begin{example}
  \label{ex:ex_NewWPIR_N3K2M2}
  Consider the same code parameters $(\const{M},\const{N},\const{K})=(2,3,2)$ as in \cref{sec:WPIR_N3K2M2}. We consider the strategy set %and the corresponding variables
\begin{IEEEeqnarray*}{c}
  \set{S}_{\textnormal{OLR}}=\bigl\{\underbrace{\bigl(\begin{smallmatrix}
      0
      \\
      1
    \end{smallmatrix}\bigr)}_{z_1},\underbrace{\bigl(\begin{smallmatrix}
      0 
      \\
      2
    \end{smallmatrix}\bigr)}_{z_2},\underbrace{\bigl(\begin{smallmatrix}
      1
      \\
      0
    \end{smallmatrix}\bigr)}_{z_3},\underbrace{\bigl(\begin{smallmatrix}
      1
      \\
      2
    \end{smallmatrix}\bigr)}_{z_4},\underbrace{\bigl(\begin{smallmatrix}
      2
      \\
      0
    \end{smallmatrix}\bigr)}_{z_5},\underbrace{\bigl(\begin{smallmatrix}
      2
      \\
      1
    \end{smallmatrix}\bigr)}_{z_6}\bigr\}.
\end{IEEEeqnarray*}
Similar to \eqref{eq:PQ_M_ZTSL_N3K2M2}, we illustrate $9$ out of the $18$ query matrices based on~\eqref{eq:queries_OLR-WPIRscheme} and the corresponding query distributions and answer lengths of the OLR MDS-WPIR scheme below:
\NiceMatrixOptions{cell-space-limits = 1mm}
\begin{IEEEeqnarray*}{c}
  \Scale[0.85]{
    \begin{pNiceMatrix}[first-row,first-col,nullify-dots]
      \small P_{\vmat{Q}_j|M}(\mat{q}_j\mid m) & \bigl(
      \begin{smallmatrix}
        0 & 0
        \\
        2 & 1
      \end{smallmatrix}\bigr) & \bigl(
      \begin{smallmatrix}
        0 & 0
        \\
        1 & 2
      \end{smallmatrix}\bigr) & \bigl(
      \begin{smallmatrix}
        2 & 1
        \\
        0 & 0
      \end{smallmatrix}\bigr) & \bigl(
      \begin{smallmatrix}
        2 & 1
        \\
        1 & 2
      \end{smallmatrix}\bigr) & \bigl(
      \begin{smallmatrix}
        1 & 2
        \\
        0 & 0
      \end{smallmatrix}\bigr) & \bigl(
      \begin{smallmatrix}
        1 & 2
        \\
        2 & 1
      \end{smallmatrix}\bigr) & \bigl(
      \begin{smallmatrix}
        1 & 0
        \\
        0 & 1
      \end{smallmatrix}\bigr) & \bigl(
      \begin{smallmatrix}
        1 & 0
        \\
        2 & 2
      \end{smallmatrix}\bigr) & \bigl(
      \begin{smallmatrix}
        0 & 1
        \\
        1 & 0
      \end{smallmatrix}\bigr) 
      \\
      1        & \frac{z_1}{3} & \frac{z_2}{3} & \frac{z_3}{3} & \frac{z_4}{3} & \frac{z_5}{3} & \frac{z_6}{3}
      & \frac{z_1}{3} & \frac{z_2}{3}  & \frac{z_3}{3}
      \\
      2        & \frac{z_2}{3} & \frac{z_1}{3} & \frac{z_5}{3} & \frac{z_6}{3} & \frac{z_3}{3} & \frac{z_4}{3}
      & \frac{z_3}{3} & \frac{z_4}{3}  & \frac{z_1}{3}
      \\*\hline
      P_{\vmat{Q}_j}(\mat{q}_j) & \frac{z_1+z_2}{3} &\frac{z_1+z_2}{6}&\frac{z_3+z_5}{3}&\frac{z_4+z_6}{6}&\frac{z_3+z_5}{6}&\frac{z_4+z_6}{6}
      & \frac{z_1+z_3}{6} & \frac{z_2+z_4}{6} & \frac{z_1+z_3}{6}
      \\*\hdottedline
      \ell_j(\mat{q}_j) & 1 & 1  & 1 & 0 & 1 & 0 & 2 & 1 & 2
      \CodeAfter
      \SubMatrix\{{1-1}{2-1}.[left-xshift=1.5em,name=C]
      \tikz \node [left] at (C-left.west) {$m$};
    \end{pNiceMatrix}}.\IEEEeqnarraynumspace\label{eq:PQ_M_new_N3K2M2}
\end{IEEEeqnarray*}

As a result, one can compute the download cost $\const{D}(\code{C}_\textnormal{OLR})$ and obtain
  \begin{IEEEeqnarray*}{c}
    \const{D}(\collect{C}_\textnormal{OLR})=2+2(z_1+z_2+z_3+z_5)\geq 2,
  \end{IEEEeqnarray*}
  which shows that $\const{R}(\collect{C}_\textnormal{OLR})$ can reach $\nicefrac{(n-k)\const{K}}{2}=1$, demonstrating a complete rate-leakage trade-off for the new MDS-WPIR scheme.
\end{example}

\section{Numerical Results}% xxx
\label{sec:submission}

Here, we compare the optimal rate-leakage trade-off curves for our three proposed MDS-WPIR schemes $\mathscr{C}_{\textnormal{ZYQT}}$, $\mathscr{C}_{\textnormal{ZTSL}}$,  and $\mathscr{C}_{\textnormal{OLR}}$. The optimal trade-off curve is obtained by solving the corresponding convex optimization problems as outlined in \eqref{eq:optimization_MaxL-download}.
% We consider the schemes' rate, subject to the optimal $\mathsf{MaxL}$ leakage metric. All curves are computed numerically using Python's CVXPY library.
For the sake of presentation, the leakage is normalized by $\log_2{\const{M}}$ bits so that its range is from $0$ to $1$.

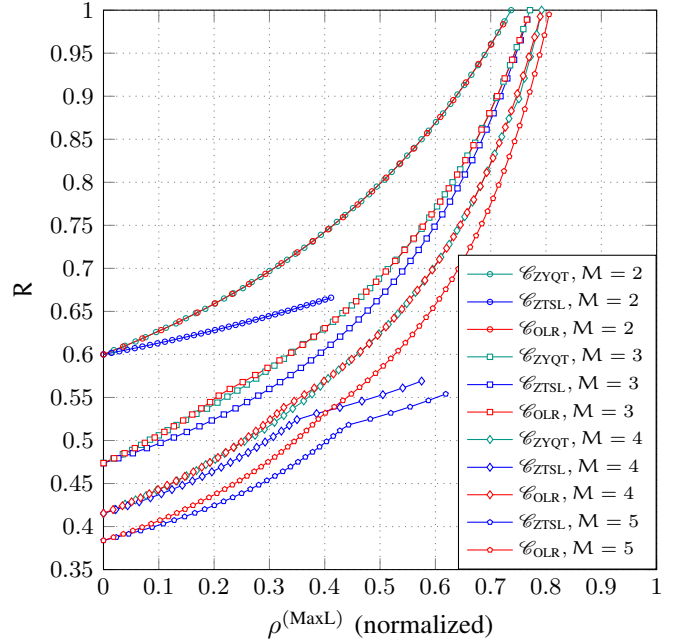
\begin{figure}[t!]
\centering
\begin{tikzpicture}%[thick, scale=0.9, every node/.style={transform shape}]
\pgfplotsset{every tick label/.append style={font=\small}}
\begin{axis}[%
width=1.0\columnwidth, % 10cm,
height=9.0cm,
at={(1.387in,0.821in)},
xmin=0,
xmax=1, % xmax=180,
xlabel={$\rho^{(\textnormal{MaxL})}$ (normalized)},
xlabel style={
	yshift=0.25ex,
	name=label},
grid style={gray,opacity=0.5,dotted},
xmajorgrids,
ymajorgrids,
% yminorgrids,
% ymode=log,
ymin=0.35, % ymin=0.0000001,
max space between ticks=20pt,
ymax=1.00,
%every axis y label/.style={at={(current axis.west)},left=5.0mm},
ylabel={$\const{R}$},
axis background/.style={fill=white},
legend cell align=left,
% no draw=none
legend style={legend style={}, font=\scriptsize, at={(axis cs: 1.0,0.35)}, anchor=south east},
]

\addplot [color=colorgreen, mark=o, mark options={solid, mark size=1pt,fill=white}]table[x=rho,y=rate] {\Figs/data/322_scheme1.txt};
\addlegendentry{$\mathscr{C}_{\textnormal{ZYQT}}, \const{M}=2$};

\addplot [color=blue, mark=o, mark options={solid, mark size=1pt,fill=white}]table[x=rho,y=rate] {\Figs/data/Rate-MaxL_minMaxL_rate_MDScodedWPIR_ZTSL_N3K2M2_v1.txt};
\addlegendentry{$\mathscr{C}_{\textnormal{ZTSL}}, \const{M}=2$};

\addplot [color=red, mark=o, mark options={solid, mark size=1pt,fill=white}]table[x=rho,y=rate] {\Figs/data/Rate-MaxL_minMaxL_rate_MDScodedWPIR_TSC_N3K2M2_v1.txt};
\addlegendentry{$\mathscr{C}_{\textnormal{OLR}}, \const{M}=2$};

\addplot [color=colorgreen, mark=square*, mark options={solid, mark size=1pt,fill=white}]table[x=rho,y=rate] {\Figs/data/323_scheme1.txt};
\addlegendentry{$\mathscr{C}_{\textnormal{ZYQT}}, \const{M}=3$};

% for verification
% \addplot [color=magenta, mark=square*, mark options={solid, mark size=1pt,fill=white}]table[x=rho,y=rate] {\Figs/data/Rate-MaxL_minMaxL_rate_MDScodedWPIR_ZYQT_N3K2M3_v1.txt};
% \addlegendentry{$\widebar{\mathscr{C}}_{\mathsf{ZYQT}}, \const{M}=3$};

\addplot [color=blue, mark=square*, mark options={solid, mark size=1pt,fill=white}]table[x=rho,y=rate] {\Figs/data/Rate-MaxL_minMaxL_rate_MDScodedWPIR_ZTSL_N3K2M3_v1.txt};
\addlegendentry{$\mathscr{C}_{\textnormal{ZTSL}}, \const{M}=3$};

\addplot [color=red, mark=square*, mark options={solid, mark size=1pt,fill=white}]table[x=rho,y=rate] {\Figs/data/Rate-MaxL_minMaxL_rate_MDScodedWPIR_TSC_N3K2M3_v1.txt};
\addlegendentry{$\mathscr{C}_{\textnormal{OLR}}, \const{M}=3$};

\addplot [color=colorgreen, mark=diamond*, mark options={solid, mark size=1.5pt,fill=white}]table[x=rho,y=rate] {\Figs/data/324_scheme1.txt};
\addlegendentry{$\mathscr{C}_{\textnormal{ZYQT}}, \const{M}=4$};

\addplot [color=blue, mark=diamond*, mark options={solid, mark size=1.5pt,fill=white}]table[x=rho,y=rate] {\Figs/data/Rate-MaxL_minMaxL_rate_MDScodedWPIR_ZTSL_N3K2M4_v1.txt};
\addlegendentry{$\mathscr{C}_{\textnormal{ZTSL}}, \const{M}=4$};

\addplot [color=red, mark=diamond*, mark options={solid, mark size=1.5pt,fill=white}]table[x=rho,y=rate] {\Figs/data/Rate-MaxL_minMaxL_rate_MDScodedWPIR_TSC_N3K2M4_v1.txt};
\addlegendentry{$\mathscr{C}_{\textnormal{OLR}}, \const{M}=4$};

\addplot [color=blue, mark=pentagon*, mark options={solid, mark size=1pt,fill=white}]table[x=rho,y=rate] {\Figs/data/Rate-MaxL_minMaxL_rate_MDScodedWPIR_ZTSL_N3K2M5_v1.txt};
\addlegendentry{$\mathscr{C}_{\textnormal{ZTSL}}, \const{M}=5$};

\addplot [color=red, mark=pentagon*, mark options={solid, mark size=1pt,fill=white}]table[x=rho,y=rate] {\Figs/data/Rate-MaxL_minMaxL_rate_MDScodedWPIR_TSC_N3K2M5_v1.txt};
\addlegendentry{$\mathscr{C}_{\textnormal{OLR}}, \const{M}=5$};

\end{axis}
\end{tikzpicture}%
%}

%\end{figure}
%%
%\end{frame}

% \end{document}
\vspace{-2ex} 
  \caption{Rate-leakage trade-off curve for the proposed MDS-WPIR protocols from $(3,2)$ MDS-coded storage with $\const{M}=2$ (circle markers), $\const{M}=3$ (square markers), $\const{M}=4$ (diamond markers), and $\const{M}=5$ (pentagon markers).}\label{fig:32}
  \vspace{-2ex} 
\end{figure}

\begin{figure}[t!]
\centering
\begin{tikzpicture}%[thick, scale=0.9, every node/.style={transform shape}]
\pgfplotsset{every tick label/.append style={font=\small}}
\begin{axis}[%
width=1.0\columnwidth, % 10cm,
height=8.25cm, % 9.0cm,
at={(1.387in,0.821in)},
xmin=0,
xmax=1, % xmax=180,
xlabel={$\rho^{(\textnormal{MaxL})}$ (normalized)},
xlabel style={
	yshift=0.25ex,
	name=label},
grid style={gray,opacity=0.5,dotted},
xmajorgrids,
ymajorgrids,
% yminorgrids,
% ymode=log,
ymin=0.5, % ymin=0.0000001,
max space between ticks=20pt,
ymax=1.00,
%every axis y label/.style={at={(current axis.west)},left=5.0mm},
ylabel={$\const{R}$},
axis background/.style={fill=white},
legend cell align=left,
% no draw=none
legend style={legend style={}, font=\scriptsize, at={(axis cs: 1,0.5)}, anchor=south east},
]

\addplot [color=colorgreen, mark=o, mark options={solid, mark size=1pt,fill=white}]table[x=rho,y=rate] {\Figs/data/532_scheme1.txt};
\addlegendentry{$\mathscr{C}_{\textnormal{ZYQT}}, \const{M}=2$};

% \addplot [color=blue, mark=square*, mark options={solid, mark size=1pt,fill=white}]table[x=rho,y=rate] {\Figs/data/532_scheme2.txt};
% \addlegendentry{$\widebar{\mathscr{C}}_{\textnormal{ZTSL}}$};

%% update the curve
\addplot [color=blue, mark=o, mark options={solid, mark size=1pt,fill=white}]table[x=rho,y=rate] {\Figs/data/Rate-MaxL_minMaxL_rate_MDScodedWPIR_ZTSL_N5K3M2_v1.txt};
\addlegendentry{$\mathscr{C}_{\textnormal{ZTSL}},\const{M}=2$};

\addplot [color=red, mark=o, mark options={solid, mark size=1pt,fill=white}]table[x=rho,y=rate] {\Figs/data/Rate-MaxL_minMaxL_rate_MDScodedWPIR_TSC_N5K3M2_v1.txt};
\addlegendentry{$\mathscr{C}_{\textnormal{OLR}}, \const{M}=2$};

\addplot [color=blue, mark=square*, mark options={solid, mark size=1pt,fill=white}]table[x=rho,y=rate] {\Figs/data/Rate-MaxL_minMaxL_rate_MDScodedWPIR_ZTSL_N5K3M3_v1.txt};
\addlegendentry{$\mathscr{C}_{\textnormal{ZTSL}},\const{M}=3$};

\addplot [color=red, mark=square*, mark options={solid, mark size=1pt,fill=white}]table[x=rho,y=rate] {\Figs/data/Rate-MaxL_minMaxL_rate_MDScodedWPIR_TSC_N5K3M3_v1.txt};
\addlegendentry{$\mathscr{C}_{\textnormal{OLR}}, \const{M}=3$};

% \addplot [color=blue, mark=diamond*, mark options={solid, mark size=1pt,fill=white}]table[x=rho,y=rate] {\Figs/data/Rate-MaxL_minMaxL_rate_MDScodedWPIR_ZTSL_N5K3M4_v1.txt};
% \addlegendentry{$\widebar{\mathscr{C}}_{\textnormal{ZTSL}},\const{M}=4$};

% \addplot [color=red, mark=10-pointed star, mark options={solid, mark size=1pt,fill=white}]table[x=rho,y=rate] {\Figs/data/Rate-MaxL_minMaxL_rate_MDScodedWPIR_TSC_N5K3M4_v1.txt};
% \addlegendentry{$\widebar{\mathscr{C}}_{\textnormal{new}}, \const{M}=4$};

\end{axis}
\end{tikzpicture}%
%}

%\end{figure}
%%
%\end{frame}

% \end{document}
\vspace{-2ex} 
  \caption{Rate-leakage trade-off curve for the proposed MDS-WPIR protocols from $(5,3)$ MDS-coded storage with $\const{M}=2$ (circle markers) and  $\const{M}=3$ (square markers).}\label{fig:53}
  \vspace{-2ex} 
\end{figure}
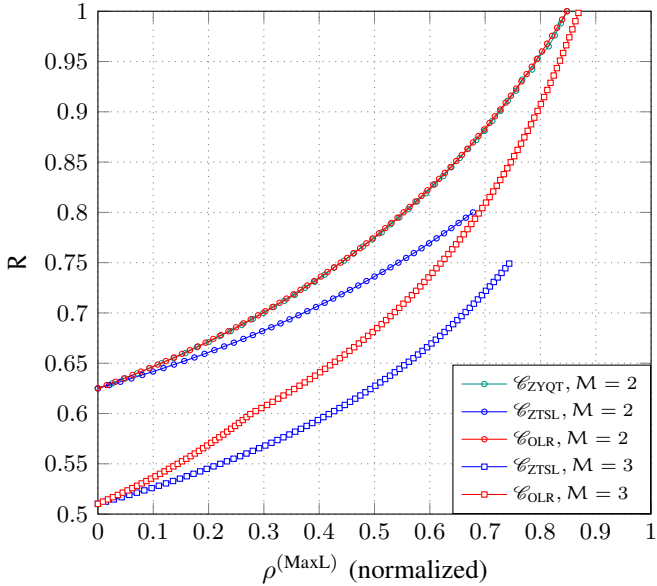

In \cref{fig:32}, we consider the case of $\const{N}=3$ servers and $\const{K}=2$, and with different number of files $\const{M}$. As can be seen from the figure by comparing the green and the blue curves, $\mathscr{C}_{\textnormal{ZYQT}}$ gives a better rate-leakage trade-off curve than $\mathscr{C}_{\textnormal{ZTSL}}$ for all considered values of $\const{M}$. Moreover, the ZTSL scheme cannot be extended to a high information leakage. On the other, the OLR scheme performs equally well as the ZYQT scheme for $\const{M}=2$ files and slightly better for a certain range of information leakage for $\const{M}=3$ and $\const{M}=4$ files, while at the same time allowing for a much smaller query space.

The corresponding rate-leakage trade-off curves for $\const{N}=5$ servers with  $\const{K}=3$ are provided in \cref{fig:53}. The same observations as in \cref{fig:32} can be made, i.e., the ZYQT scheme outperforms the ZTSL scheme, while the proposed OLR scheme yields an equal trade-off curve as the ZYQT scheme for $\const{M}=2$ files. As the query space is significant for the ZYQT scheme for $\const{M}=3$ files, we were not able to solve the corresponding convex optimization problem as outlined in \eqref{eq:optimization_MaxL-download} and therefore no curve for $\const{M} > 2$ is presented. However, as mentioned previously, a nice feature of the OLR scheme is its smaller  query space, and hence the corresponding optimization problem in \eqref{eq:optimization_MaxL-download} can be readily solved even for $\const{M} = 3$. In particular, we have $\ecard{\set{S}_{\textnormal{ZYQT}}}=216000>\ecard{\set{S}_{\textnormal{OLR}}}=1500$ for $\const{M}=3$.

% performs the best for $\const{M}=2$, while for $\const{M}=3$, $\Bar{\mathscr{C}}_{\textnormal{ZTSL}}$ performs the best. For $\const{M}=4$, $\Bar{\mathscr{C}}_{\textnormal{ZTSL}}$ performs better for small leakages, while there is a crossing with the curve for $\Bar{\mathscr{C}}_{\textnormal{ZYQT}}$ when the leakage becomes larger (the crossing is for a leakage around $0.4$).$

% Interestingly, $\Bar{\mathscr{C}}_{\mathsf{ZTSL}}$ cannot achieve a higher rate than
%$\nicefrac{L}{D_{\mathsf{max}}} = \nicefrac{4}{5}$. 

\section{Conclusion}
\label{sec:conclusion}

This work is the first to consider WPIR for coded storage. In particular, we  proposed and compared three WPIR protocols for the case where the data is encoded by an MDS code and stored across multiple servers. Allowing for some leakage on the identity of the requested file index  allows for a higher download rate, and we showed that the optimal trade-off of download rate and information leakage using the MaxL privacy metric is the solution to a convex optimization problem for all three proposed protocols. %

% We conclude by pointing out that on the last page the columns need to
% be balanced. Instructions for that purpose are given in the source
% file (they are commented out).

%%%%%%
%% To balance the columns at the last page of the paper use this
%% command somewhere at the top of the first column of the last page:
%%
% \enlargethispage{-5cm} 
%%
%% where the exact amount of page reduction has to be adapted to the
%% actual situation.
%%
%% If the balancing should occur in the middle of the references, use
%% the following trigger:
%%
% \IEEEtriggeratref{3}
%%
%% which triggers a \newpage (i.e., new column) just before the given
%% reference number. Note that you need to adapt this if you modify
%% the paper. The "triggered" command can be changed if desired:
%%
% \IEEEtriggercmd{\enlargethispage{-20cm}}
%%
%%%%%%

%%%%%%
%% References:
%% We recommend the usage of BibTeX:
%%
\bibliographystyle{IEEEtran}
\bibliography{defshort1,biblioHY}

\begin{thebibliography}{10}
\providecommand{\url}[1]{#1}
\csname url@samestyle\endcsname
\providecommand{\newblock}{\relax}
\providecommand{\bibinfo}[2]{#2}
\providecommand{\BIBentrySTDinterwordspacing}{\spaceskip=0pt\relax}
\providecommand{\BIBentryALTinterwordstretchfactor}{4}
\providecommand{\BIBentryALTinterwordspacing}{\spaceskip=\fontdimen2\font plus
\BIBentryALTinterwordstretchfactor\fontdimen3\font minus
  \fontdimen4\font\relax}
\providecommand{\BIBforeignlanguage}[2]{{%
\expandafter\ifx\csname l@#1\endcsname\relax
\typeout{** WARNING: IEEEtran.bst: No hyphenation pattern has been}%
\typeout{** loaded for the language `#1'. Using the pattern for}%
\typeout{** the default language instead.}%
\else
\language=\csname l@#1\endcsname
\fi
#2}}
\providecommand{\BIBdecl}{\relax}
\BIBdecl

\bibitem{ChorGoldreichKushilevitzSudan95_1}
B.~Chor, O.~Goldreich, E.~Kushilevitz, and M.~Sudan, ``Private information
  retrieval,'' in \emph{Proc. 36th Annu. IEEE Symp. Found. Comp. Sci. (FOCS)},
  Milwaukee, WI, USA, Oct. 23--25, 1995, pp. 41--50.

\bibitem{ChorGoldreichKushilevitzSudan98_1}
------, ``Private information retrieval,'' \emph{J. ACM}, vol.~45, no.~6, pp.
  965--982, Nov. 1998.

\bibitem{BeimelIshaiKushilevitRaymond02_1}
A.~Beimel, Y.~Ishai, E.~Kushilevitz, and J.-F. Raymond, ``Breaking the
  {$O(n^{1/(2k-1)})$} barrier for information-theoretic private information
  retrieval,'' in \emph{Proc. 43rd Annu. IEEE Symp. Found. Comp. Sci. (FOCS)},
  Vancouver, BC, Canada, Nov. 16--19, 2002, pp. 261--270.

\bibitem{Yekhanin10_1}
S.~Yekhanin, ``Private information retrieval,'' \emph{Commun. ACM}, vol.~53,
  no.~4, pp. 68--73, Apr. 2010.

\bibitem{CorriganGibbsKogan20_1}
H.~Corrigan-Gibbs and D.~Kogan, ``Private information retrieval with sublinear
  online time,'' in \emph{Proc. 39th Annu. Int. Conf. Theory Appl. Crypto.
  Techn. (EUROCRYPT)}, Zagreb, Croatia, May 10--14, 2020, pp. 44--75.

\bibitem{ChanHoYamamoto15_1}
T.~H. Chan, S.-W. Ho, and H.~Yamamoto, ``Private information retrieval for
  coded storage,'' in \emph{Proc. IEEE Int. Symp. Inf. Theory (ISIT)}, Hong
  Kong, China, Jun. 14--19, 2015, pp. 2842--2846.

\bibitem{Freij-HollantiGnilkeHollantiKarpuk17_1}
R.~Freij-Hollanti, O.~W. Gnilke, C.~Hollanti, and D.~A. Karpuk, ``Private
  information retrieval from coded databases with colluding servers,''
  \emph{SIAM J. Appl. Algebra Geom.}, vol.~1, no.~1, pp. 647--664, Nov. 2017.

\bibitem{TajeddineGnilkeElRouayheb18_1}
R.~Tajeddine, O.~W. Gnilke, and S.~El~Rouayheb, ``Private information retrieval
  from {MDS} coded data in distributed storage systems,'' \emph{IEEE Trans.
  Inf. Theory}, vol.~64, no.~11, pp. 7081--7093, Nov. 2018.

\bibitem{SunJafar17_1}
H.~Sun and S.~A. Jafar, ``The capacity of private information retrieval,''
  \emph{IEEE Trans. Inf. Theory}, vol.~63, no.~7, pp. 4075--4088, Jul. 2017.

\bibitem{BanawanUlukus18_1}
K.~Banawan and S.~Ulukus, ``The capacity of private information retrieval from
  coded databases,'' \emph{IEEE Trans. Inf. Theory}, vol.~64, no.~3, pp.
  1945--1956, Mar. 2018.

\bibitem{KumarLinRosnesGraellAmat19_1}
S.~Kumar, H.-Y. Lin, E.~Rosnes, and A.~Graell~i Amat, ``Achieving maximum
  distance separable private information retrieval capacity with linear
  codes,'' \emph{IEEE Trans. Inf. Theory}, vol.~65, no.~7, pp. 4243--4273, Jul.
  2019.

\bibitem{LinKumarRosnesGraellAmat18_2}
H.-Y. Lin, S.~Kumar, E.~Rosnes, and A.~Graell~i Amat, ``Asymmetry helps:
  Improved private information retrieval protocols for distributed storage,''
  in \emph{Proc. IEEE Inf. Theory Workshop (ITW)}, Guangzhou, China, Nov.
  25--29, 2018.

\bibitem{LinKumarRosnesGraellAmatYaakobi19_1}
H.-Y. Lin, S.~Kumar, E.~Rosnes, A.~Graell~i Amat, and E.~Yaakobi,
  ``Weakly-private information retrieval,'' in \emph{Proc. IEEE Int. Symp. Inf.
  Theory (ISIT)}, Paris, France, Jul. 7--12, 2019, pp. 1257--1261.

\bibitem{SamyTandonLazos19_1}
I.~Samy, R.~Tandon, and L.~Lazos, ``On the capacity of leaky private
  information retrieval,'' in \emph{Proc. IEEE Int. Symp. Inf. Theory (ISIT)},
  Paris, France, Jul. 7--12, 2019, pp. 1262--1266.

\bibitem{LinKumarRosnesGraellAmatYaakobi21_1}
H.-Y. Lin, S.~Kumar, E.~Rosnes, A.~Graell~i Amat, and E.~Yaakobi, ``The
  capacity of single-server weakly-private information retrieval,'' \emph{IEEE
  J. Sel. Areas Inf. Theory}, vol.~2, no.~1, pp. 415--427, Mar. 2021.

\bibitem{LinKumarRosnesGraellAmatYaakobi22_1}
------, ``Multi-server weakly-private information retrieval,'' \emph{IEEE
  Trans. Inf. Theory}, vol.~68, no.~2, pp. 1197--1219, Feb. 2022.

\bibitem{QianZhouTianLiu22_1}
C.~Qian, R.~Zhou, C.~Tian, and T.~Liu, ``Improved weakly private information
  retrieval codes,'' in \emph{Proc. IEEE Int. Symp. Inf. Theory (ISIT)}, Espoo,
  Finland, Jun. 26--Jul. 1, 2022, pp. 2840--2845.

\bibitem{SamyAttiaTandonLazos21_1}
I.~Samy, M.~Attia, R.~Tandon, and L.~Lazos, ``Asymmetric leaky private
  information retrieval,'' \emph{IEEE Trans. Inf. Theory}, vol.~67, no.~8, pp.
  5352--5369, Aug. 2021.

\bibitem{ZhouGuoTian20_1}
R.~Zhou, T.~Guo, and C.~Tian, ``Weakly private information retrieval under the
  maximal leakage metric,'' in \emph{Proc. IEEE Int. Symp. Inf. Theory (ISIT)},
  Los Angeles, CA, USA, Jun. 21--26, 2020, pp. 1089--1094.

\bibitem{YakimenkaLinRosnesKliewer22_1}
Y.~Yakimenka, H.-Y. Lin, E.~Rosnes, and J.~Kliewer, ``Optimal
  rate-distortion-leakage tradeoff for single-server information retrieval,''
  \emph{IEEE J. Sel. Areas Commun.}, vol.~40, no.~3, pp. 832--846, Mar. 2022.

\bibitem{WengYakimenkaLinRosnesKliewer22_1}
C.-W. Weng, Y.~Yakimenka, H.-Y. Lin, E.~Rosnes, and J.~Kliewer, ``Generative
  adversarial user privacy in lossy single-server information retrieval,''
  \emph{IEEE Trans. Inf. Forens. Secur.}, vol.~17, pp. 3495--3510, 2022.

\bibitem{Smith09_1}
G.~Smith, ``On the foundations of quantitative information flow,'' in
  \emph{Proc. 12th Int. Conf. Found. Softw. Sci. Comput. Struct. (FoSSaCS)},
  York, U.K., Mar. 22--29, 2009, pp. 288--302.

\bibitem{BartheKopf11_1}
G.~Barthe and B.~K{\"o}pf, ``Information-theoretic bounds for differentially
  private mechanisms,'' in \emph{Proc. 24th IEEE Comput. Secur. Found. Symp.
  (CSF)}, Cernay-la-Ville, France, Jun. 27--29, 2011, pp. 191--204.

\bibitem{IssaWagnerKamath20_1}
I.~Issa, A.~B. Wagner, and S.~Kamath, ``An operational approach to information
  leakage,'' \emph{IEEE Trans. Inf. Theory}, vol.~66, no.~3, pp. 1625--1657,
  Mar. 2020.

\bibitem{ZhuYanQiTang20_1}
J.~Zhu, Q.~Yan, C.~Qi, and X.~Tang, ``A new capacity-achieving private
  information retrieval scheme with (almost) optimal file length for coded
  servers,'' \emph{IEEE Trans. Inf. Forens. Secur.}, vol.~15, pp. 1248--1260,
  2020.

\bibitem{ZhouTianSunLiu20_1}
R.~Zhou, C.~Tian, H.~Sun, and T.~Liu, ``Capacity-achieving private information
  retrieval codes from {MDS}-coded databases with minimum message size,''
  \emph{IEEE Trans. Inf. Theory}, vol.~66, no.~8, pp. 4904--4916, Aug. 2020.

\end{thebibliography}
%%
%% where we here have assume the existence of the files
%% definitions.bib and bibliofile.bib.
%% BibTeX documentation can be obtained at:
%% http://www.ctan.org/tex-archive/biblio/bibtex/contrib/doc/
%%%%%%
%% Or you use manual references (pay attention to consistency and the
%% formatting style!):

\end{document}